\documentclass[doublecol]{epl2}

\title{$\eta$-mesic nuclei in relativistic mean-field theory}
\shorttitle{$\eta$ mesic nuclei in relativistic mean-field theory} 

\author{C.\ Y.\ Song\inst{1} \and X.\ H.\ Zhong\inst{2} \and L.\
Li\inst{1} \and P.\ Z.\ Ning \inst{1}} \shortauthor{C.\ Y.\ Song
\etal}

\institute{
  \inst{1} Department of Physics, Nankai University, Tianjin 300071, China\\
  \inst{2} Institute of High Energy Physics,
       Chinese Academy of Sciences, Beijing 100039, China
}  {Second
pacs description}

\abstract{ With the $\eta$-nucleon ($\eta N$) interaction Lagrangian
deduced from chiral perturbation theory, we study the possible
$\eta$-mesic nuclei in the framework of relativistic mean-field
theory. The $\eta$ single-particle energies are sensitive to the
$\eta N$ scattering length, and increase monotonically with the
nucleon number $A$. If the scattering length is in the range of
$a^{\eta\mathrm{N}}=0.75\sim1.05$ fm and the imaginary potential
$V_{0}\sim15$ MeV, some discrete states of $_{\eta}^{12}$C,
$_{\eta}^{16}$O and $_{\eta}^{20}$Ne should be identified in
experiments. However, when the scattering length
$a^{\eta\mathrm{N}}< 0.5 $ fm, or the imaginary potential $V_{0} >
30$ MeV, no discrete $\eta$ meson bound states could be observed in
experiments.}

\begin{document}

\maketitle

\section{Introduction}
Since the $\eta$-mesic nuclei were  predicted by Haider et al.,
\cite{b1,b2}  the topics on the $\eta N$ interactions and
$\eta$-mesic nuclei are studied extensively.
Although all of the theory models predict that the interaction
between $\eta$-meson and nucleon is attractive, its strength (i.e.
the predicted $\eta$ nuclear potential ) has strong model
dependence, spans from about -20  MeV to -100 MeV
\cite{b3,op,efm,op1}.

Because of the uncertainties of the $\eta$ nuclear potentials, the
predictions of the $\eta$-mesic nuclei are very different in
different models
\cite{b1,b4,b6,b7,b8,b9,b10,b11,b12,b13,b14,b15,Trho,b761}. For
example, some models predicted that $\eta$-mesic nuclei could be
found in the nuclei with nucleon number $A>10$ \cite{b1}, while
some other models predicted that they could be found in very light
nuclei with $A\geq 2$ \cite{b13,b14,b15}.

Experimentally, several experiments bad been performed
\cite{b29,b30}, but no evidence of $\eta$-mesic nuclei was found.
Recently, Sokol et al. \cite{b31} claimed that they observed a
$\eta$-mesic nucleus, $^{11}_{\eta}$C, by measuring the invariant
mass of correlated $\pi^{+}n$ pairs in a photo-mesonic reaction. And
more recently, M. Pfeiffer {\sl et al}.\ \cite{b32} also claimed
they observed some information of a $\eta$-mesic nucleus,
$^{3}_{\eta}$He. To get a further understanding on $\eta$-mesic
nuclei, more studies, both in theory and experiments, are needed.

In our previous work, the $\eta N$ interaction Lagrangian had been
derived from the chiral perturbation theory (ChPT) \cite{zxh}, in
which  the off shell term has been related with the $\eta N$
scattering length by a off-shell term parameter $\kappa$. Combining
this $\eta N$  Lagrangian with the Lagrangian for nucleons in
relativistic mean field theory (RMF), we have obtained the equations
of motion for nucleons and mesons. By solving the these equations
self-consistently in RMF, the static properties of $\eta$-mesic
nuclei, such as the single-particle energy spectra, are gotten.
Similar method can be found in the study of kaonic nuclei as well
\cite{ddd,zxh1}. In the RMF calculations, with the existing data of
the scattering lengths, the lower limits of the 1s state
single-particle $\eta$ binding energies are $9\pm7$ MeV, and the
upper limits are $70\pm 10$ MeV. With large scattering length
$a^{\eta\mathrm{N}}=0.75\sim1.05$ fm and small imaginary potential
$V_{0}\sim 15$ MeV, the discrete bound states of $_{\eta}^{12}$C,
$_{\eta}^{16}$O and $_{\eta}^{20}$Ne may be identified in
experiments.

This work is organized as follows. In the subsequent section, the
Lagrangian density is given, the equations of motion for nucleons
and the meson fields $\sigma$, $\omega$, $\rho$, and $\eta$ are
deduced, the imaginary part of the self-energies are introduced.
We then present our results and discussions in Sec. III. Finally a
summary is given in Sec. IV.

\section{Framework}

\subsection{Lagrangian and equations of motion}

In relativistic mean field theory, the standard Lagrangian density
for an ordinary nucleus can be written as \cite{d1,d2}
\begin{eqnarray}
{\mathcal{L}}_{0} ={\mathcal{L}}_{\mathrm{Dirac}}
 +{\mathcal{L}}_{\sigma}
 +{\mathcal{L}}_{\omega}
 +{\mathcal{L}}_{\rho}
 +{\mathcal{L}}_{A},
\end{eqnarray}
where
\begin{eqnarray}
{\mathcal{L}}_{\mathrm{Dirac}} &=&
 \bar{\Psi}_N(i\gamma^{\mu}\partial_{\mu}-M_{N})\Psi_N, \\
{\mathcal{L}}_{\sigma} &=&
 \frac{1}{2}\partial_{\mu}\sigma\partial^{\mu}\sigma
 -\frac{1}{2}m_{\sigma}^2\sigma^2
 -g_{\sigma N }\bar{\Psi}_N\sigma\Psi_N
\nonumber\\
& &
 -\frac{1}{3}g_{2}\sigma^3
 -\frac{1}{4}g_{3}\sigma^4,\\
{\mathcal{L}}_{\omega} &=&
 -\frac{1}{4}F_{\mu\nu}F^{\mu\nu}
 +\frac{1}{2}m_{\omega}^2\omega_{\mu}\omega^{\mu}
\nonumber\\
& &
 -g_{\omega N}\bar{\Psi}_N\gamma^{\mu}\Psi_N\omega_{\mu}, \\
{\mathcal{L}}_{\rho} &=&
 -\frac{1}{4}\vec{G}_{\mu\nu}\vec{G}^{\mu\nu}
 +\frac{1}{2}m_{\rho}^2
\vec{\rho}_{\mu}\cdot\vec{\rho}^{\mu}\nonumber\\
& & -g_{\rho N}\bar{\Psi}_{N}\vec{\rho}^{\mu}\cdot \vec{I}\Psi_{N},\\
{\mathcal{L}}_{A} &=&
  -\frac{1}{4}H_{\mu\nu}H^{\mu\nu}
  -e\bar{\Psi}_N\gamma_{\mu}I_{c}A^{\mu}\Psi_N,
\end{eqnarray}
with
\begin{eqnarray}
F_{\mu\nu} &=&
 \partial_{\nu}\omega_{\mu}-\partial_{\mu}\omega_{\nu},\\
\vec{G}_{\mu\nu} &=&
 \partial_{\nu}\vec{\rho}_{\mu}-\partial_{\mu}\vec{\rho}_{\nu}, \\
H_{\mu\nu} &=& \partial_{\nu}A_{\mu}-\partial_{\mu}A_{\nu}.
\end{eqnarray}
In the above equations, the meson fields are denoted by $\sigma$,
$\omega_{\mu}$, and $ \rho_{\mu}$, with masses $m_{\sigma}$,
$m_{\omega}$, $m_{\rho}$, respectively.  $\Psi_N$ is the nucleon
field with corresponding mass $M_N$. $A_{\mu}$ is the
electromagnetic field. $g_{\sigma N} $, $g_{\omega N} $, and
$g_{\rho N} $ are, respectively, the $\sigma$-$N$, $\omega$-$N$,
and $\rho$-$N$ coupling constants. $I_{c}=(1+\tau_{3})/2$ is the
Coulomb interaction operator with $\tau_3$ being the third
component of the isospin Pauli matrices for nucleons. $I$ is the
nucleon isospin operator. In the calculations, we adopt the NL-SH
parameter set (see Tab. \ref{paramtab}) \cite{d3}, which describes
the properties of finite nuclei reasonably.

For an $\eta$-nucleus system, another Lagrangian density
${\mathcal{L}}_{\mathrm{\eta}}$ describing the $\eta N$
interactions should be added to ${\mathcal{L}}_0$. In this work,
the Lagrangian density ${\mathcal{L}}_{\mathrm{\eta}}$ is adopted
the one deduced from the heavy baryon chiral perturbation theory
up to the next-to-leading-order terms \cite{zxh}, which is given
by
\begin{eqnarray} \label{teffL}
\mathcal{L_{\eta }}  &=&
 \frac{1}{2}\partial^{\mu}\eta\partial_{\mu}\eta
 -\frac{1}{2}\left(
  m_{\eta}^2
-\frac{\Sigma_{\eta N}}{f_{\pi}^2}\bar{\Psi}_{N}\Psi_{N}
 \right) \eta^2\nonumber\\
 &&+\frac{1}{2}\cdot\frac{\kappa}{f_{\pi}^2}\bar{\Psi}_{N}\Psi_{N}\partial^{\mu}\eta\partial_{\mu}\eta,
\end{eqnarray}
where $m_{\eta}=547.311$ MeV corresponds to the mass of
$\eta$-meson, $\Sigma_{\eta\mathrm{N}}$ is the $\eta\mbox{N}$
sigma term, $\kappa$ is a parameter of the ``off-shell" term.
$f_{\pi}\simeq93$ MeV is the pseudoscalar meson decay constants.
According to our previous work \cite{zxh}, we set $\Sigma_{\eta
N}=280$ MeV. The ``off-shell" term parameter $\kappa$ was
determined by the $\eta\mbox{N}$ scattering length
$a^{\eta\mathrm{N}}$,
 \begin{eqnarray} \label{akap}
\kappa =4\pi f_{\pi}^2
 \left(
  \frac{1}{m_{\eta}^2}+\frac{1}{m_{\eta}M_{N}}
 \right)
a^{\eta N} -\frac{\Sigma_{\eta N}}{m_{\eta}^2}.
\end{eqnarray}
The scattering length has large uncertainties, which scatters in a
large range $a^{\eta\mathrm{N}}=0.2\sim 1.1$ fm
\cite{Green,Renard,Arndt,Kaiser,IJMPA}. Thus, the corresponding
value of $\kappa$ is in the range of $(-0.13 \sim 0.40)$ fm. It
should be emphasized that in the ChPT the contributions of
$N^*(1535)$ can not be seen directly, however, its contributions are
included by the scattering length, which relates to the resonance
$N^*(1535)$ directly.

\begin{table}[ht]
\begin{center}
\caption{Parameters used in the present calculations.}
\label{paramtab}
\begin{tabular}{|ccccc|c|c|c|}\hline\hline\\
$M_N$ & $m_{\sigma}$ & $m_{\omega}$ & $m_{\rho}$ & \\
\\
939.0 & 526.059 &783.0 & 763.0 &\\
\hline
\\
$g_{\sigma N} $ & $g_{\omega N} $ & $g_{\rho N} $   & $g_3$& $g_2$ \\
\\
10.444  & 12.945  & 8.766  & -15.8337 & -6.9099 $fm^{-1}$\\
\hline\hline
\end{tabular}
\end{center}
\end{table}

In the mean field approximation, the meson-fields $\sigma$,
$\omega_{\mu}$, and $\rho_{\mu}$, and the photons $A_{\mu}$ are
replaced with their mean values, $\langle \sigma \rangle$,
$\langle \omega_{\mu}\rangle$, $\langle\rho_{\mu}\rangle$ and
$\langle A_{\mu}\rangle$, respectively. For a spherical nucleus,
only the mean values of the time components $\langle
\omega_{0}\rangle$, $\langle\rho_{0}\rangle$ and $\langle
A_{0}\rangle$ remain, which are denoted by $\omega_{0}$, and
$\rho_{0}$, and $A_{0}$ respectively. From the Lagrangian for the
$\eta$-nucleus system, the equations of motion for nucleons,
$\omega$, $\sigma$, $\rho$, and photons are deduced, which are
given by
\begin{eqnarray}
& \left\{\vec{\alpha}\cdot\vec{P}+\beta [M_{N}+S(r)
]+V(r)\right\}\Psi_{N}=\mathcal{E} \Psi_{N},
&\label{eossimp1}\\
&\left(-\nabla^2+m_{\sigma}^2\right)\sigma_{0}= -g_{\sigma N}
\bar{\Psi}_{N}\Psi_{N} -g_{2}\sigma_{0}^2-g_{3}\sigma_{0}^3,
&\label{eossimp2}\\
& \left(-\nabla^2+m_{\omega}^2\right)\omega_{0}=g_{\omega N}
\bar{\Psi}_{N}\gamma^{0} \Psi_{N},
&  \label{eossimp3}\\
& \left(-\nabla^2+m_{\rho}^2\right)\rho_{0}=g_{\rho N}
\bar{\Psi}_{N}\gamma^{0}I \Psi_{N} ,
& \label{eossimp4}\\
& -\nabla^2A_0=e \bar{\Psi}_{N}\gamma^{0}I_{c} \Psi_{N}, &
\label{eossimp5}
\end{eqnarray}
with
\begin{eqnarray}
S(r)&=&g_{\sigma N}\sigma_{0}-\frac{1}{2}\cdot\frac{\Sigma_{\eta
N}}{f_{\pi}^2}\eta^{2}-\frac{1}{2}\cdot\frac{\kappa}{f_{\pi}^2}\partial^{\mu}\eta\partial_{\mu}\eta,\label{s}\\
V(r)&=&g_{\omega N}\omega_{0}+g_{\rho N}\tau_3\rho_{0}+e
I_{c}A_{0}\label{v}.
\end{eqnarray}
In the calculation the spacial terms of the last term in
Eq.(\ref{s}) are neglected for a simplicity.
And the equation of motion for $\eta$ meson is derived as
\begin{eqnarray}
\bigg[-\nabla^{2}+(m_{\eta}^{2}-E^{2})+\Pi\bigg]\eta=0,
\end{eqnarray}
with
\begin{eqnarray}
\Pi=-\frac{1}{f_{\pi}^{2}(1+\frac{\kappa}{f_{\pi}^{2}}\varrho_{s})}
(\kappa m_{\eta}^{2} +\Sigma_{\eta N}) \varrho_{s}.
\end{eqnarray}
In the above equations, $\mathcal{E}$ is the nucleon single-particle
energy, $E$ is the single-particle energy for $\eta$ meson,
$\varrho_{s}=\bar{\Psi}_{N}\Psi_{N}$ is the scalar density of
nucleons, and $\Pi$ is the self-energy of the $\eta$ meson in the
nucleus.

\subsection{Imaginary potential}

Within the framework of RMF model, there is only a real part for
the self-energy of the $\eta$ meson in the nucleus. Considering
there are strong absorption for the $\eta$-mesons in a nucleus, in
the realistic calculations the imaginary part of the self-energy
should be included. Thus, as done in Refs. \cite{zxh1,ddd}, we
assume a specific form for the self-energy:
\begin{eqnarray}
\widetilde{\Pi}&=&\Pi+i\left[-2(\mathrm{Re} E)f V_{0}
\frac{\varrho}{\varrho_0}\right].
\end{eqnarray}
The imaginary part of the potential $\mathrm{Im}U$ is adopted the
simple ``$t\varrho$" form, namely, $\mathrm{Im}U=-f
V_{0}\varrho/\varrho_0$. $f$ is a suppression factor, which will be
discussed later. $V_{0}$ is the imaginary potential depth at normal
nuclear density $\varrho_0$, which has strong model dependence. The
shallowest value of $V_{0}\sim10$ MeV is given by fitting larger
scattering length using the ``$t\varrho$" form \cite{b15}. While
Waas and Weise studied the s-wave interactions of $\eta$-meson in
nuclear medium, and got $V_{0}\simeq 22$ MeV \cite{b3}. Inoue and
Oset also obtained $V_{0}\simeq 29$ MeV with chiral unitary approach
\cite{op1}. Using the chiral doublet model to incorporate the medium
effects of the $N^*(1535)$ resonance, Jido and Nagahiro \emph{et
al.} predicted the largest imaginary potential depth $V_{0}\simeq
50$ MeV \cite{b6,b12,b761}. Chiang {\sl et al}.\ \cite{op} suggested
the imaginary potential depth in the range of $(12\sim49)$ MeV by
assuming that the mass of the $N^*(1535)$ did not change in the
medium. Thus, in the present work, we set the imaginary potential
depth $V_{0}$ in the range of $10\sim 50$ MeV to cover all the
possible ranges.

Considering the decay channels should be reduced for the
$\eta$-meson being bound in a nucleus, the suppression factor,
$f$, is introduced to multiply the imaginary part to decrease the
imaginary potentials (widths)\footnote{The energy of a free system
is larger than the energy of a bound system, thus, for a decay
channel its phase space should be suppressed for a bound system.
As an example, we can see Eq.(\ref{f1}) and Eq.(\ref{f2}).}. This
method has been used to calculate the width of kaonic nuclei
\cite{ddd,Galaa,zxh1}. There are two main decay channels for
$\eta$-mesic nuclei. One is the mesonic decay channel, $\eta
N\rightarrow \pi N$. The corresponding suppression factor is given
by \cite{ddd,Galaa,zxh1}
\begin{eqnarray}\label{f1}
f_1=\frac{M_{01}^3}{M_1^3}\sqrt{\frac{[M_1^2-M_{+}^2][M_1^2-M_{-}^2]}
{[M_{01}^2-M_{+}^2][M_{01}^2-M_{-}^2]}} \Theta(M_1-M_+),
\end{eqnarray}
where $M_{01}=m_\eta+M_N$, $M_{+}=m_{\pi}+M_{N}$,
$M_{-}=M_{N}-m_{\pi}$ and $M_{1}=\mathrm{Re}E+M_N$ is the energy of
the bound system $\eta N$. The other channel is the non-mesonic
decay channel, $\eta NN\rightarrow NN$, and the corresponding
suppression factor is \cite{ddd,Galaa,zxh1}
\begin{eqnarray} \label{f2}
f_2=\frac{M_{02}^3}{M_2^3}\sqrt{\frac{[M_2^2-4M_{N}^2]M_2^2}
{[M_{02}^2-4M_{N}^2]M_{02}^2}}\Theta (M_2-2M_{N}),
\end{eqnarray}
where $M_{02}=m_\eta+2M_N$, $M_{2}=\mathrm{Re}E+2M_N$ correspond to
the energies of the free system and the bound system of $\eta NN$,
respectively. The mesonic decay and non-mesonic decay are studied in
Ref. \cite{0412036}, the ratio for the two decay modes are about
90\% and 10\%, respectively. Thus the suppression factor $f$ can be
written as
\begin{eqnarray}
f=0.9 f_1+0.1 f_2.
\end{eqnarray}

\begin{table}
\caption{The single-particle $\eta$ binding energies,
$B^{s,p}_{\eta}=m_\eta-Re E$  and the widths, $\Gamma$, (both in
MeV), in various nuclei for $\kappa$=-0.13 fm ($a^{\eta N}=0.20$
fm), where the complex eigenenergies are,
$E=-B^{s,p}_\eta+m_\eta-i\Gamma/2$.} \label{t.lbl}
\begin{center}
\begin{tabular}{cccccccc} \hline \hline\\
&&\multicolumn{2}{c}{\underline{$V_{0}=15$}}&\multicolumn{2}{c}{\underline{$V_{0}=30$}}&\multicolumn{2}{c|}{\underline{$V_{0}=50$}}\\
&&$B_{\eta}^{s,p}$&$\Gamma$&
  $B_{\eta}^{s,p}$&$\Gamma$&
  $B_{\eta}^{s,p}$&$\Gamma$\\ \hline

\\
$^{ 16}_{\eta} $O &$1s$& \textmd{-   } &      & \textmd{  -  } &     & \textmd{ -   } &      \\
\\
$^{ 20}_{\eta} $Ne&$1s$& \textbf{4.1 } &21.2  & \textmd{  -  } &     & \textmd{ -   } &      \\
\\
$^{ 24}_{\eta} $Mg&$1s$& \textbf{6.1 } &23.8  & \textbf{2.8 } &51.3 & \textmd{ -   } &    \\
\\
$^{ 28}_{\eta} $Si&$1s$& \textbf{7.9 } &26.0  & \textbf{4.9 } &55.3 & \textmd{ -   } &    \\
\\
$^{ 32}_{\eta} $S &$1s$& \textbf{8.5 } &26.5  & \textbf{5.3 } &56.0 & \textmd{  -  } &    \\
\\
$^{ 36}_{\eta} $Ar&$1s$& \textbf{8.9 } &25.8  & \textbf{6.1 } &54.3 & \textbf{1.8 } &95.3\\
\\
$^{ 40}_{\eta} $Ca&$1s$& \textbf{9.2 } &25.4  & \textbf{6.8 } &53.1 & \textbf{3.2 } &92.7\\
\\
$^{ 44}_{\eta} $Ti&$1s$& \textbf{10.0} &25.8  & \textbf{7.7 } &53.7 & \textbf{4.4 } &93.2\\
\\
\\
$^{132}_{\eta} $Xe&$1s$& \textbf{15.2} &27.8  & \textbf{14.2} &55.9 & \textbf{12.1} &94.0\\
                  &$1p$& 7.3  &26.6  & 5.8 &54.1 & 3.0  &92.1\\
$^{208}_{\eta} $Pb&$1s$& \textbf{16.3} &28.4  & \textbf{15.6} &56.8 & \textbf{13.9} &94.6\\
                  &$1p$& 9.7  &28.4  & 8.9  &56.9 & 7.1  &95.0\\
  \hline \hline

\end{tabular}
\end{center}
\end{table}

\begin{table}
\caption{The single-particle $\eta$ binding energies,
$B^{s,p}_{\eta}=m_\eta-Re E$  and the widths, $\Gamma$, (both in
MeV), in various nuclei for $\kappa$=0.04 fm ($a^{\eta N}=0.50$ fm),
where the complex eigenenergies are,
$E=-B^{s,p}_\eta+m_\eta-i\Gamma/2$.} \label{t1.lbl}
\begin{center}
\begin{tabular}{cccccccc} \hline \hline\\
&&\multicolumn{2}{c}{\underline{$V_{0}=15$}}&\multicolumn{2}{c}{\underline{$V_{0}=30$}}&\multicolumn{2}{c|}{\underline{$V_{0}=50$}}\\
&&$B_{\eta}^{s,p}$&$\Gamma$&
  $B_{\eta}^{s,p}$&$\Gamma$&
  $B_{\eta}^{s,p}$&$\Gamma$\\ \hline\\
$^{12}_{\eta}$C&$1s$ & \textbf{26.2} &30.7  & \textbf{23.2} &64.0& \textbf{17.2} &109.9\\
               &$1p$ & \textmd{-   } &      & \textmd{-   } &    & \textmd{-   } &     \\
  \\
$^{16}_{\eta}$O&$1s$ & \textbf{24.8} &27.2  & \textbf{22.9} &55.3& \textbf{18.9} &94.5 \\
               &$1p$ & \textmd{-   }  &     & \textmd{-   } &    & \textmd{-   } &     \\
 \\
$^{20}_{\eta}$Ne&$1s$& \textbf{27.5} &26.6  & \textbf{25.8} &53.8& \textbf{22.6} &91.5 \\
               &$1p$ & \textmd{-   } &      & \textmd{-   } &    & \textmd{ -  } &     \\
  \\
$^{24}_{\eta}$Mg&$1s$& \textbf{31.2} &28.5  & \textbf{29.6} &57.4& \textbf{26.5} &97.1 \\
                &$1p$& \textbf{8.8 } &22.2  & \textbf{5.8 } &46.7& \textmd{  - } &     \\
  \\
  \\
$^{28}_{\eta}$Si&$1s$& \textbf{34.5} &29.5  & \textbf{33.1} &59.4& \textbf{30.1} &100.1\\
               &$1p$ & \textbf{13.1} &25.4  & \textbf{10.4} &52.3& \textbf{5.4 } &92.2 \\
  \\
  \\
$^{32}_{\eta}$S &$1s$& \textbf{36.1} &30.6  & \textbf{34.4} &61.6& \textbf{30.9} &103.7\\
                &$1p$& \textbf{13.4} &24.3  & \textbf{10.8} &49.9& \textbf{5.9 } &86.9 \\
  \\
  \\
$^{36}_{\eta}$Ar&$1s$& \textbf{35.5} &29.3  & \textbf{34.0} &59.0& \textbf{30.9} &99.3 \\
                &$1p$& \textbf{15.0} &24.2  & \textbf{12.8} &49.6& \textbf{8.5 } &85.9 \\
  \\
  \\
$^{40}_{\eta}$Ca&$1s$& \textbf{35.1} &28.6  & \textbf{33.8} &57.5& \textbf{31.1} &96.8 \\
                &$1p$& \textbf{16.8} &24.8  & \textbf{14.7} &50.7& \textbf{10.9} &87.3 \\
  \\
  \\
$^{44}_{\eta}$Ti&$1s$& \textbf{36.0} &28.0  & \textbf{34.8} &56.9& \textbf{32.4} &95.5 \\
                &$1p$& \textbf{18.8} &25.7  & \textbf{16.9} &51.7& \textbf{13.4} &88.7 \\
  \\
 \hline \hline

\end{tabular}
\end{center}
\end{table}

\subsection{Single-particle $\eta$
binding energy and width}

Then the modified Klein-Gordon equation can be expressed as,
\begin{eqnarray}\label{mdf}
\bigg[-\nabla^{2}+(m_{\eta}^{2}-E^{2})+\widetilde{\Pi}\bigg]\eta=0.
\end{eqnarray}
The complex eigenenergy is
\begin{eqnarray}
E=-B^{s,p}_\eta+m_\eta-i\Gamma/2,
\end{eqnarray}
where the real part corresponds to the single-particle $\eta$
binding energy, which is defined as
\begin{eqnarray}
B^{s,p}_{\mathrm{\eta}}=m_{\mathrm{\eta}}-\mathrm{Re}E,
\end{eqnarray}
and the imaginary part of the complex eigenenergy corresponds to
the width
\begin{eqnarray}
\Gamma=-2\mathrm{Im} E.
\end{eqnarray}

Solving the equations (\ref{eossimp1}) --- (\ref{eossimp5}) and
Eq.(\ref{mdf}) self-consistently, we can obtain the
single-particle energy spectra and widths of $\eta$ mesic nuclei.

\section{Results and discussions}

In this section, the single-particle energy spectra and the widths
of the possible $\eta$-mesic nuclei, such as $^{12}_{\eta}$C,
$^{16}_{\eta}$O, $^{20}_{\eta}$Ne, $^{24}_{\eta}$Mg,
$^{28}_{\eta}$Si, $^{32}_{\eta}$S, $^{36}_{\eta}$Ar,
$^{40}_{\eta}$Ca and $^{44}_{\eta}$Ti are calculated in RMF. For the
uncertainties of the parameter $\kappa$ (i.e. the scattering length
$a^{\eta\mathrm{N}}$), which give large uncertainties for the $\eta
$ nuclear potentials, we choose four values of $\kappa$ ($-0.13$,
$0.04$, $0.19$ and $0.40$ fm corresponding to $a^{\eta N}=0.20$,
$0.50$, $0.75$, $1.05$ fm) to cover all the possible scattering
lengths. In each case, we also suppose $V_{0}=15$, $30$ and $50$
MeV, respectively, which can cover all the possible ranges of the
imaginary potential. The results, including the single-particle
$\eta$ binding energies ($B^{s,p}_{\eta}$) and the widths
($\Gamma$), for $\kappa=-0.13$ fm ($a^{\eta N}=0.20$ fm) and
$\kappa=0.04$ fm ($a^{\eta N}=0.50$ fm) are shown in Tab.\
\ref{t.lbl} and Tab.\ \ref{t1.lbl}, respectively. And the results
for $\kappa=0.19$ , $0.40$ fm ($a^{\eta N}=0.75$ , $1.05$ fm) are
listed in Tab.\ \ref{Calll}.

For $a^{\eta N}=0.20$ fm (see Tab.\ \ref{t.lbl}), it is found that
the imaginary potential depth $V_{0}$ has effects on the lighter
nuclei to form $\eta$ quasi-bound states. For example, with
$V_{0}=15$ MeV, quasi-bound states can be found with nucleon number
$A\geq20$, however, they are only found in the $A\geq36$ nuclei with
$V_{0}=50$ MeV. The 1s state single-particle binding energies are
$(9\pm7)$ MeV, increasing with the nucleon number. The widths are
much larger than the single-particle binding energies even we use
the smallest $V_{0}=15$ MeV. Thus, no $\eta$-mesic nuclei can be
observed in experiments.

For $a^{\eta N}=0.50$ fm (see Tab.\ \ref{t1.lbl}) the ground state
single-particle binding energies are ($26\pm10$) MeV. If the
imaginary part $V_{0}=15$ MeV, the decay widths are comparable with
the the binding energies, thus, in this case the $\eta$-mesic nuclei
maybe observed in the light nuclei when $a^{\eta N}\geq 0.50$ fm. On
the contrary, when $a^{\eta N}<0.50$ fm, no $\eta$-mesic nuclei can
be observed in experiments.

\begin{center}
\begin{largetable}
 \caption{The single-particle $\eta$ binding energies, $B^{s,p}_{\eta}=m_\eta-Re E$
 and the widths, $\Gamma$, (both in MeV), in various nuclei for $\kappa$=0.19 fm
($a^{\eta N}=0.75$ fm) and $\kappa$=0.40 fm ($a^{\eta N}=1.05$ fm),
where the  complex eigenenergies are,
$E=-B^{s,p}_\eta+m_\eta-i\Gamma/2$.}
 \label{Calll}
\begin{tabular}{cccccccc|cccccc}  \hline \hline\\
&&\multicolumn{6}{c|}{$\kappa$=0.19 fm($a^{\eta N}=0.75$ fm)}&\multicolumn{6}{c}{$\kappa$=0.40 fm($a^{\eta N}=1.05$ fm)}\\
\hline \hline\\
 &&\multicolumn{2}{c}{\underline{$V_{0}=15$}}&\multicolumn{2}{c}{\underline{$V_{0}=30$}}&\multicolumn{2}{c|}{\underline{$V_{0}=50$}}
 &\multicolumn{2}{c}{\underline{$V_{0}=15$}}&\multicolumn{2}{c}{\underline{$V_{0}=30$}}&\multicolumn{2}{c}{\underline{$V_{0}=50$}}
 \\ 
&&$B_{\eta}^{s,p}$&$\Gamma$&
  $B_{\eta}^{s,p}$&$\Gamma$&
  $B_{\eta}^{s,p}$&$\Gamma$&
  $B_{\eta}^{s,p}$&$\Gamma$&
  $B_{\eta}^{s,p}$&$\Gamma$&
  $B_{\eta}^{s,p}$&$\Gamma$
\\ \hline\\
$^{12}_{\eta}$C&$1s$ & \textbf{46.2} &34.9& \textbf{43.8} &70.3& \textbf{38.6} &118.2& \textbf{69.8} &35.5  & \textbf{67.8} &70.9 & \textbf{63.3} &118.2\\
               &$1p$ & \textbf{6.6 } &19.4& \textbf{3.2 } &40.5&       -       &  -  & \textbf{23.4} &23.7  & \textbf{21.1} &47.8 & \textbf{15.7} &83.3 \\
 \\
 \\
$^{16}_{\eta}$O&$1s$ & \textbf{43.2} &29.6& \textbf{41.6} &59.8& \textbf{38.2} &100.8& \textbf{65.4} &31.4  & \textbf{64.1} &62.9 & \textbf{61.0} &104.9\\
               &$1p$ & \textbf{13.2} &21.0& \textbf{10.8} &43.0& \textbf{6.0 } &75.0 & \textbf{31.1} &24.4  & \textbf{29.3} &49.2 & \textbf{25.6} &83.1 \\
  \\
  \\
$^{20}_{\eta}$Ne&$1s$& \textbf{46.3} &27.6& \textbf{45.1} &55.3& \textbf{42.4} &92.7 & \textbf{69.1} &26.9  & \textbf{68.1} &53.8 & \textbf{66.0} &89.8 \\
               &$1p$ & \textbf{18.6} &22.5& \textbf{16.6} &45.7& \textbf{12.2} &80.4 & \textbf{37.8} &23.5  & \textbf{36.5} &47.2 & \textbf{33.4} &81.4 \\
  \\
  \\
$^{24}_{\eta}$Mg&$1s$& \textbf{50.9} &28.9& \textbf{49.8} &57.9& \textbf{47.2} &96.9 & \textbf{74.7} &28.5  & \textbf{73.7} &57.0 & \textbf{71.5} &95.0 \\
                &$1p$& \textbf{25.2} &24.8& \textbf{23.4} &50.7& \textbf{19.5} &87.4 & \textbf{46.0} &26.2  & \textbf{44.7} &52.6 & \textbf{41.8} &88.3 \\
  \\
  \\
$^{28}_{\eta}$Si&$1s$& \textbf{55.1} &30.5& \textbf{53.9} &61.2& \textbf{51.3} &102.2& \textbf{79.7} &30.1  & \textbf{78.7} &60.3 & \textbf{76.4} &100.3\\
               &$1p$ & \textbf{31.0} &27.4& \textbf{29.3} &55.4& \textbf{25.6} &93.7 & \textbf{53.1} &28.7  & \textbf{51.8} &57.5 & \textbf{48.8} &96.3 \\
  \\
$^{32}_{\eta}$S &$1s$& \textbf{57.4} &32.2& \textbf{56.1} &63.7& \textbf{53.0} &106.4& \textbf{82.9} &31.9  & \textbf{81.7} &63.8 & \textbf{78.8} &106.1\\
               &$1p$ & \textbf{30.9} &26.3& \textbf{29.2} &53.0& \textbf{25.6} &89.7 & \textbf{52.8} &27.2  & \textbf{51.5} &54.6 & \textbf{48.6} &91.3 \\
  \\
  \\
$^{36}_{\eta}$Ar&$1s$& \textbf{56.2} &30.7& \textbf{55.0} &61.6& \textbf{52.2} &102.8& \textbf{81.2} &29.6  & \textbf{80.1} &59.1 & \textbf{77.8} &98.3 \\
               &$1p$ & \textbf{32.5} &26.2& \textbf{30.9} &52.7& \textbf{27.6} &89.0 & \textbf{54.3} &26.6  & \textbf{53.1} &53.3 & \textbf{50.5} &89.1 \\
 \\
 \\
$^{40}_{\eta}$Ca&$1s$& \textbf{55.4} &29.5& \textbf{54.3} &59.0& \textbf{51.8} &98.5 & \textbf{79.8} &28.3  & \textbf{78.9} &56.5 & \textbf{76.8} &94.0 \\
               &$1p$ & \textbf{34.3} &26.1& \textbf{33.0} &52.6& \textbf{30.1} &88.7 & \textbf{56.2} &26.4  & \textbf{55.2} &52.8 & \textbf{52.9} &88.3 \\
  \\
  \\
$^{44}_{\eta}$Ti&$1s$& \textbf{56.3} &28.7& \textbf{55.3} &58.1& \textbf{53.1} &97.0 & \textbf{80.7} &28.1  & \textbf{79.9} &56.3 & \textbf{78.0} &93.6 \\
               &$1p$ & \textbf{36.8} &26.8& \textbf{35.5} &53.3& \textbf{32.8} &89.6 & \textbf{59.0} &26.9  & \textbf{58.1} &54.0 & \textbf{55.8} &90.1 \\
  \\
\hline \hline
\end{tabular}
\end{largetable}
\end{center}

For $a^{\eta N}=(0.75\sim1.05)$ fm (see Tab.\ \ref{Calll}), the 1s
state single-particle binding energies are in the range of $(48\pm
10\sim 70\pm 10)$ MeV, and those of 1p states are in the region of
$(15\pm12 \sim 38\pm 21)$ MeV, increasing monotonically with the
nucleon number $A$. The separations of the single-particle $\eta$
binding energies between the 1p and 1s states are on the magnitude
of $(30\pm10\sim 35\pm 13)$ MeV, decreasing with the increment of
the nucleon number in general. When $V_{0}\sim15$ MeV, the sum of
the half widths of the 1s and 1p states are narrower than the
separations of the single-particle $\eta$ binding energies between
1s and 1p states for C, O, Ne, which implies that some discrete
states should be identified in experiments for these nuclei.
However, if $V_{0}>30$ MeV no $\eta$ mesic nuclei could be observed
in experiments according to our calculations.

From Tab.\ \ref{t1.lbl} and Tab.\ \ref{Calll}, it is found that
the widths of the 1s states are in the ranges of $(28\pm 7\sim
104\pm 14)$ MeV and those of 1p states are $(26\pm 3\sim 89\pm 7)$
MeV, respectively, for $V_{0}=(15\sim 50)$ MeV. The imaginary
potential depth $V_{0}$ has slight effects on the values of the
single-particle energy $B^{s,p}_{\eta}$, the effects decrease with
the increment of $V_{0}$. For example, if we change $V_{0}$ from
15 MeV to 50 MeV, the single-particle energies decrease about
$(3\sim 8)$ MeV for both 1s and 1p states .

\section{Summary}

Some possible $\eta$ mesic nuclei from $^{12}_{\eta}$C to
$^{44}_{\eta}$Ti have been studied in RMF. The  $\eta$
single-particle energy is sensitive to the $\eta N$ scattering
length (i.e. ``off-shell" term parameter $\kappa$). In the whole
possible range for the scattering length, the lower limits of the 1s
state single-particle $\eta$ binding energies are $(9\pm7)$ MeV, and
the upper limits are $(70\pm 10)$ MeV. The widths of 1s states are
in the ranges of $(28\pm 7\sim 104\pm 14)$ MeV and those of 1p
states are $(26\pm 3\sim 89\pm 7)$ MeV.

When the scattering length $a^{\eta N}=(0.75\sim 1.05)$ fm, and the
imaginary potential $V_{0} \leq 15$ MeV, the sum of the half widths
of the 1s and 1p states for $^{12}_{\eta}$C, $^{16}_{\eta}$O and
$^{20}_{\eta}$Ne are smaller than the separations of the
single-particle binding energies between the two low-lying states of
these $\eta$-mesic nuclei, which implies that discrete $\eta$ meson
bound states may be identified in experiments in these nuclei.
However, when the scattering length $a^{\eta N}< 0.5 $ fm, or the
imaginary potential $V_{0} > 30$ MeV, no discrete $\eta$ meson bound
states could be identified in experiments.

Finally, we should point out that it is an attempt to study the
$\eta$ mesic nuclei with the $\eta N$ interaction deduced from ChPT.
In our method the contributions of resonances, such as $N^*(1535)$,
are only included indirectly by the $\eta N$ scattering length,
which relates to the resonances directly. The imaginary potential is
phenomenologically introduced in this paper, which has a large
uncertainty. Thus, more realistic $\eta N$ interaction which
introduces the resonances naturally, and more fundamental imaginary
potential should be pursued in the future work.

\acknowledgments The authors thank N. Auerbach for many good
suggestions. This work was supported, in part, by the Natural
Science Foundation of China (grants 10575054 and 10775145), China
Postdoctoral Science Foundation, and K. C. Wong Education
Foundation, Hong Kong.

\end{document}